\documentclass[epj]{svjour}

\usepackage{graphicx}
\usepackage{fancyhdr}

\def\mytitle{My title}
\def\myauthors{My name}
\def\mytype{My type of session}
\def\mysession{My session}

\def\mytitle{R-Hadron and long lived particle searches at the LHC} 
\def\myauthors{S. Bressler}    
\def\mytype{Contributed Talk}
\def\mysession{Colliders - SUSY Phenomenology}

\begin{document}
\title{R-Hadron and long lived particle searches at the LHC}
\subtitle{}
\author{S. Bressler\inst{1}, on behalf of the ATLAS and CMS collaborations
\
 \
}                     
%
%
\institute{Technion - Israel Institute of Technology
}
%
\date{}
\abstract{
If long lived charged particles exist, and produced at the LHC, they may travel with velocity significantly slower than the speed of light. This unique signature was not considered during the design of the LHC experiments, ATLAS and CMS. As a result, hardware and trigger capabilities need to be evaluated.\\
Model independent approaches for finding long lived particles with the LHC experiments are introduced. They are tested using two bench marks, one in GMSB and one in Split SUSY. The focus is on hardware and trigger issues, as well as reconstruction methods developed by ATLAS and CMS.\\
Both experiments suggest time of flight (TOF) based methods. However, the implementation is different. In ATLAS a first $\beta$ estimation is done already at the trigger level. CMS also uses dE/dx to estimate $\beta$.
\PACS{
      {SUSY}{}   \and
      {LHC}{} \and
      {GMSB}{} \and
      {R-Hadron}{}
     } 
} 

\maketitle
%

\section{Introduction}
\label{intro}
Long lived charged particles are allowed by many models of physics beyond the standard model (SM). In general, different models have different characteristics. However, long lived charged particle have a common characteristic in many models; when produced at the LHC some of them would travel with velocity significantly lower than the speed of light. This unique signature makes the search for it a model independent search.\\
The LHC experiments, ATLAS and CMS, were designed to fully exploit the LHC discovery potential and make new discoveries. Nevertheless, the scenario of slowly moving particles was not considered. This fact makes the search for it a non conventional and exciting challenge.\\
The second section of this paper introduces two model contexts with long lived charged particles. The third section describes the unique signature of such particles in the LHC. Trigger considerations are discussed in section four, and discovery methods are introduced in section five.

\section{model context}
\label{sec:1}

The original motivation for the search for long lived charged particles arises from Gauge mediated SUSY breaking (GMSB) models. In GMSB the lightest SUSY particle (LSP) is the gravitino. If R-Parity is conserved, the next to lightest SUSY particle (NLSP) must decay weakly to the LSP and hence may become long lived.\\
The GMSB particle spectrum is determined by a set of five parameters: $\Lambda$- the SUSY breaking scale, $M_m$- the messenger mass scale, $N_m$- the number of messenger fields, $tan\beta$- the ratio of the vacuum expectation values of the two Higgs fields, $sign(\mu)$- the sign of the $\mu$ term and $C_{grav}$- the scale factor of the gravitino mass, which determines the NLSP lifetime ($\tau_{NLSP}$ $\sim$ $C_{grav}^2$).\\
Choosing, for example, $\Lambda$=30TeV , $M_m$=250TeV, $N_m$ =3, $tan\beta$=5, $sign(\mu)$=+ and $C_{grav}$=5000  (ATLAS's GMSB5) results in 23pb production cross section of long lived slepton, stau or selectron, with mass $M_{stau}$=102.2GeV and $M_{selectron}$=100.3GeV.\\
If R-Parity is assumed, in GMSB5, any event must contain 2 sleptons, each one accompanied by a lepton (tau or electron respectively). Since the two NLSPs are produced at the end of cascade decays, they are produced with different $p_T$. Figure 1 shows the $\beta$ spectrum of the muons and NLSPs in GMSB5. As can be seen, the $\beta$ spectrum of the NLSPs covers a large range, with a tendency towards the high $\beta$ values.\\
%
\begin{figure}[htbp]
\label{fig:1}       
%
\centerline{
\includegraphics[width=0.45\textwidth,height=0.3\textwidth,angle=0]{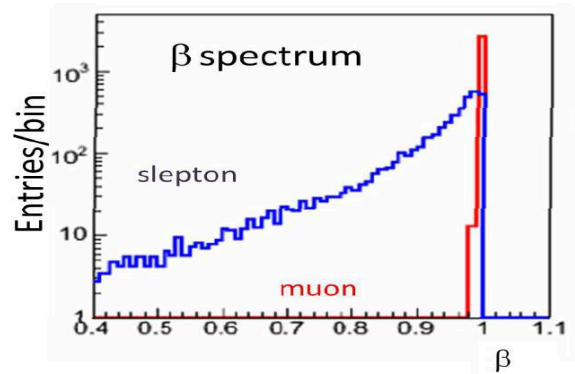}
}
\caption{The $\beta$ spectrum the GMSB5 long lived sleptons}
\end{figure}
Split SUSY may also yield long lived charged particle \cite{RefATLAS1}. In split SUSY the gluino NLSP hadronizes into long lived R-Hadrons. Unlike GMSB, the two gluinos are produced at the primary interaction, so that both R-Hadrons are expected to have the same $p_T$ and $\beta$. The production cross section of R-Hadron is as high as several nano-barns for low mass gluinos but decreases logarithmically.  The $\beta$ spectrum gets softer with increasing mass (Figure 2).\\
%
\begin{figure}[htbp]
\label{fig:2}       
\centerline{
\includegraphics[width=0.45\textwidth,height=0.4\textwidth,angle=0]{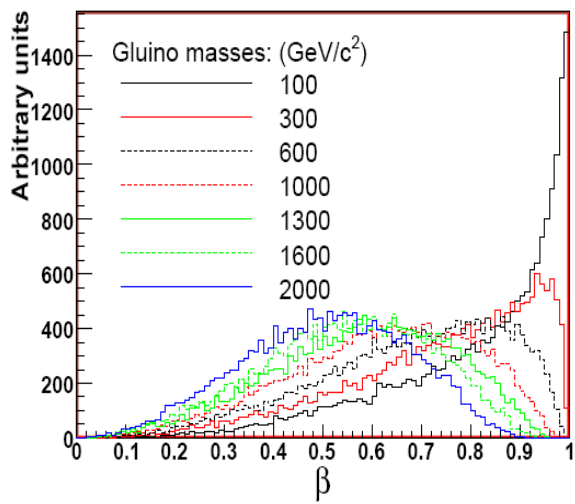}
}
\caption{$\beta$ distribution for different gluino mass}
\end{figure}
\section{Signature}
\label{sec:2}
A long lived charge particle interacts in the detector like a heavy muon. Neither EM shower nor substantial hadronic showers are expected. Therefore, the signal to look for at the LHC is a charged particle with low $\beta$ which reaches the muon chambers. The mass of a particle with $\beta$$\sim$1 can not be measured since it looks just like high $p_T$ muon.\\
When searching for slow particles at the LHC, it is essential to pay attention to the detector dimensions. The length of ATLAS, for example, is over 20m, and since the collision period at the LHC is 25ns, three events coexist in the detector at the same time.\\
To match correctly event fragments from different sub-detectors, bunch crossing identification (BCID) is crucial. BCID is based on time measurement, each detector is calibrated with respect to particles which move almost at the speed of light. When ${\beta}$$<$$1$ hits may be marked with a wrong BCID and lost during data taking. \\
For example, when $\beta$$=$$0.7(0.6)$ the efficiency to be in the correct BC decreases below to 80\%(20\%) \cite{RefATLAS2}. Therefore, in order to be able to measure low $\beta$ particle more events have to be read out.
\section{Trigger}
\label{sec:3}
Both ATLAS and CMS have a trigger system to reduce the data taking rate from 40MHz to $\sim$100Hz. The first level trigger selection is done by custom HW and the high level trigger is done by dedicated SW. An event must pass all trigger levels in order to be read out.\\
A long lived charged particle is most likely to trigger as a muon. However, different models may result in different trigger scenarios. In GMSB5 the two sleptons are produced at the end of different cascade decays, hence they are produced with different $\beta$. According to figure 1 at least one slepton is likely to have high enough $\beta$ to trigger in the correct BC.\\
In Split SUSY, the gluinos are directly produced. Therefore both of the R-Hadrons may be slow and the trigger may be on the wrong BC. As a result, inner detector information may be lost in the previous BC, so that a matching requirement between muon spectrometer and inner detector track may cause loss of events. If the R-hadron is produced neutral and changes charge in the calorimeter, there is no inner detector track even in the previous BC.\\
\section{Discovery methods}
\label{sec:4}
Discovery of a new long-lived particle means measuring unknown mass. The mass is reconstructed from $\beta$ and momentum measurements. Different methods using measurements of time of flight (TOF) or specific energy loss (dE/dx) are used for $\beta$ measurements in the different experiments. In the following we concentrate on the TOF based discovery methods suggested by ATLAS and dE/dx based method from CMS. Details on the CMS TOF methods are included in another contribution to these proceedings.\\
\subsection{Discovery methods using the ATLAS detector}
\label{sec:5}
Typical event flow in ATLAS starts with a collision, where the particle is created. The particle then propagates through the detector, leaving hits along its path. The hits transform into electronic signals, which contain the position and time information. In the next step the signals are used by the trigger system to make the event selection.\\
Events that pass all the trigger levels are processed through common reconstruction algorithms that are not accessible by the users. During reconstruction the signals are combined into tracks which are written out for analysis. The individual signal information, specifically the time information, is not written out and so is not available at later stages. The last step is the analysis, where private algorithms use the tracks to discover the physics. We will show that in the case of long lived charged particle the discovery can not await the analysis step. Moreover, we will show that in the barrel, significant work can be done already at the second level of the trigger.\\
\subsubsection{TOF and mass measurement at the ATLAS trigger level 2 }
\label{sec:6}
The muon barrel trigger chambers (RPC) of ATLAS have a time resolution of 3.125ns. TOF calculation was added to the barrel level 2 algorithm muFast to get initial estimation of the particle's speed \cite{RefATLAS2}.\\
Figures 3 shows that the measured $\beta\prime$s are within 5\% of the simulated $\beta\prime$s. The resolution at the trigger level 2 is less than 5\% for all $\beta\prime$s, and slow particles are distinguished from particles which move at the speed of light. Despite the accurate $\beta$ measurement, when using the expected rate of hight $p_T$ muons, the background from high $p_T$ muons overwhelms the signal in $\beta$. \\
%
\begin{figure}[htbp]
\label{fig:3}       
\centerline{
\includegraphics[width=0.45\textwidth,height=0.4\textwidth,angle=0]{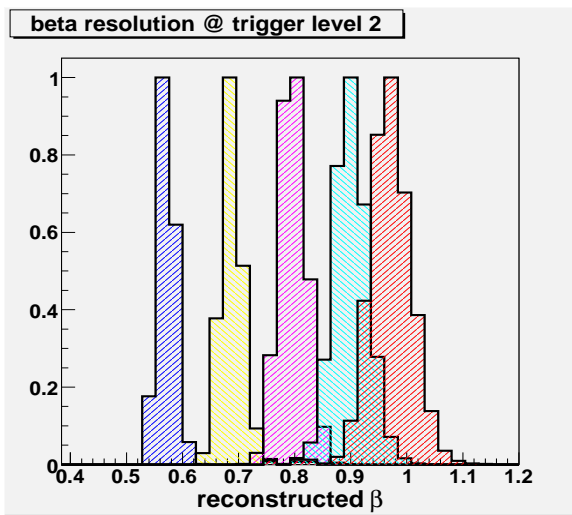}
}
\caption{Reconstructed $\beta$: for slepton with $\beta$=0.6,0.7,0.8,0.9 and muon $\beta$=1. Mean values are 0.573,0.688,0.796,0.899 and 0.97 respectively. }
\end{figure}
For that reason a slepton hypothesis based on mass measurement was added. Figure 4 shows that for slepton hypothesis with the cuts $p_T$$>$$40GeV$, $m_{slepton}$$>$$40GeV$ and $\beta$$<$$0.97$, the signal is seperated well from the background.
%
\begin{figure}[htbp]
\label{fig:4}       
\centerline{
\includegraphics[width=0.45\textwidth,height=0.4\textwidth,angle=0]{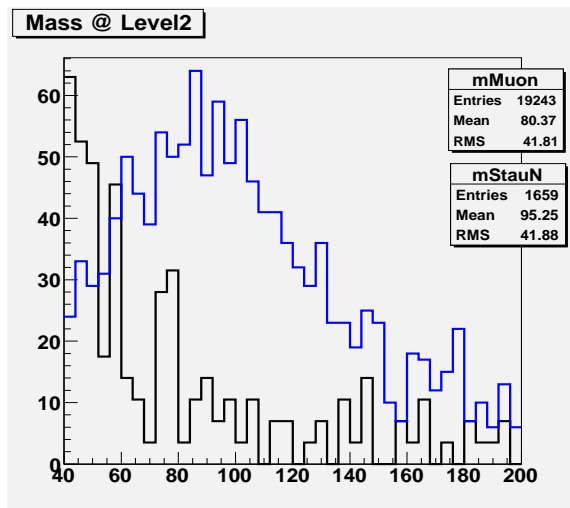}
}
\caption{Reconstructed mass: GMSB5 signal compare to expected rate of high $p_T$ muons. For slepton hypothesis with the cuts $p_T$$>$$40GeV$, $m_{slepton}$$>$$40GeV$ and $\beta$$<$$0.97$, the signal is seperated well from the background.}
\end{figure}
\subsubsection{A $\beta$ reconstruction algorithm at ATLAS}
\label{sec:6}
When a charged particle passes through the ATLAS monitored drift tubes detector (MDT) it leaves clusters of ionized atoms. The electrons drift to the wire in the center of the tube. The radius from which the electron drifts to the wire is calculated from a time measurement of the drift time, $r=r(t_{drift})$. An MDT chamber consists of 6 or 8 layers of tubes in two multilayers. The segment in each chamber is tangent to the radii. During segment reconstruction some noise hits are ignored.\\
The drift time, $t_{drift}$, relies on knowing the TOF to each tube which is calibrated with respect to particles which travel at the speed of light. In the case of slow particles the calculated drift time is larger than the true drift time, $t_{drift}$ = $t_{drift true}$ + $\Delta$t. Here $\Delta$t corresponds to the delay of the slow particle with respect to particle which moves at the speed of light. Therefore, the reconstructed radii are larger than the real radii. Larger radii result in badly fitted segment or wrong direction of the segment.\\
The $\beta$ reconstruction algorithm relies on the long time window of the MDT and BCID from the inner detector \cite{RefATLAS2}. The base of the algorithm is a loop over possible $\Delta$t`s. In each iteration, the MDT digit times, and hence radii, are changed, and a segment is created from the re-timed digits. The TOF is estimated from the $\Delta$t that minimizes the $\chi^2$ of the segments. Together with the information from the segments in the trigger chambers the mass distribution presented in figure 5 is obtained.
%
\begin{figure}[htbp]
\label{fig:5}       
\centerline{
\includegraphics[width=0.45\textwidth,height=0.4\textwidth,angle=0]{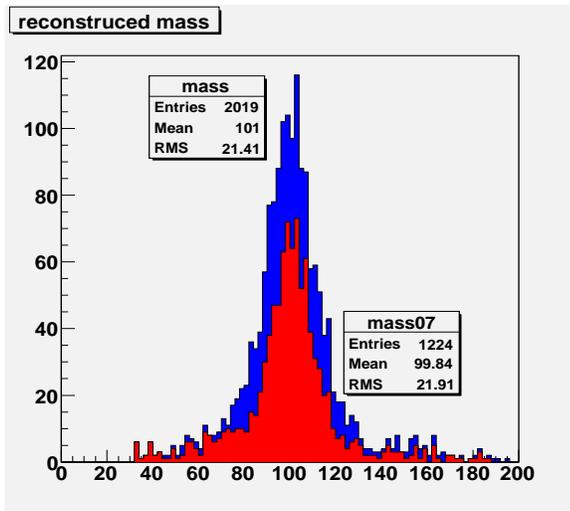}
}
\caption{Mass resolution using ATLAS's reconstruction algorithm}
\end{figure}
\subsubsection{A second $\beta$ reconstruction algorithm at ATLAS}
\label{sec:7}
A second TOF based algorithm uses the MDT chambers in a similar manner. However, it starts from a large sample of sparticles, thus it was done for R-Hadrons \cite{RefATLAS3}.\\
After achieving a large sample of R-Hadrons, the sample is divided in momentum bins. In each bin, the $\beta$ which  minimizing the average $\chi^2$  is chosen . Then $\beta$ is fitted as a function of the momentum $\beta(p)$. Finally the mass is calculated m=p/$\gamma(p)\beta(p)$. Figure 6 presents the reconstructed mass.\\
%
\begin{figure}[htbp]
\label{fig:6}       
\centerline{
\includegraphics[width=0.45\textwidth,height=0.4\textwidth,angle=0]{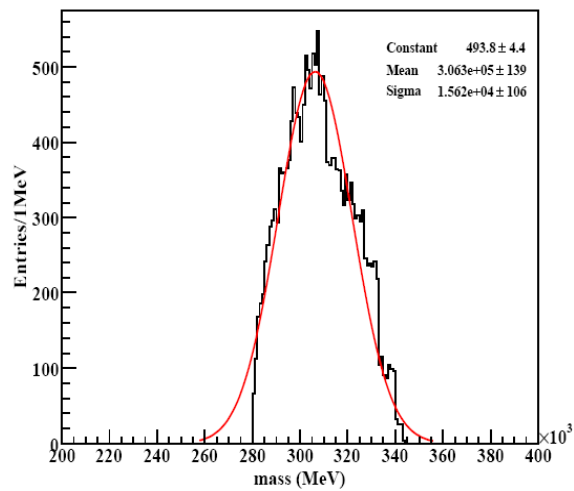}
}
\caption{Mass measurement as function of momentum}
\end{figure}
\subsection{$\beta$ reconstruction at CMS}
\label{sec:8}
As mention earlier, details on the CMS TOF method are included in another contribution to these proceedings, so only CMS reconstruction based on dEdx is described below.\\
\subsubsection{Tracker $\beta$ measurement with dE/dx at CMS}
\label{sec:9}
The expected energy deposition of a particle travelling through matter (dE/dx) is a function of $\beta$, so can provide a $\beta$ measurement.\\
The CMS silicon tracker provides a measure of the energy deposition per module. The energy deposition is calibrated using SM particles. Additional optimization is done in order to minimize effects caused by the late arrival of the slow particle to the module. Finally, an average specific energy loss, dE/dx, per track is calculated.\\
A fit of the Bethe-Block formula to the observed energy deposition as a function of particle momentum is used to estimate the particle's velocity \cite{RefCMS}. Figure 7 shows the $\beta$ resolution using the dE/dX measurement at the CMS trackers.
%
\begin{figure}[htbp]
\label{fig:7}       
\centerline{
\includegraphics[width=0.45\textwidth,height=0.4\textwidth,angle=0]{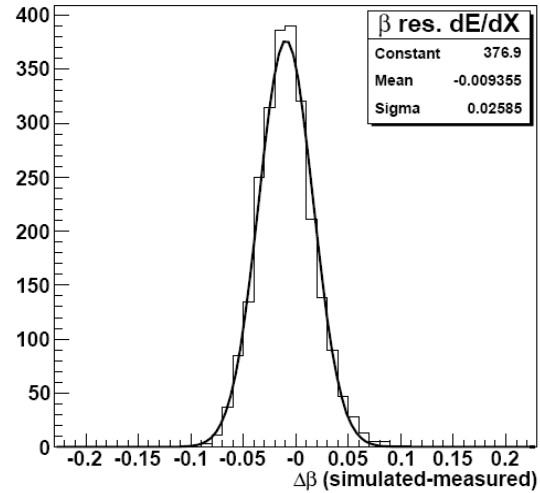}
}
\caption{$\beta$ resolution using the dE/dX measurement at CMS}
\end{figure}
\section{Conclusions}
\label{sec:10}
If long lived charged particles exist within the TeV scale, the LHC experiments, ATLAS and CMS, are capable of discovering it. However, this requires paying attention and modifying pre-envisioned details of detector and trigger operation. In particular, reading out data of additional BCs will increase efficiency for the lower $\beta$ range. It was also shown that the search for long lived charged particle can not await the analysis stage, but it must be started in the trigger and reconstruction stages.\\
%
%
%
%

\begin{thebibliography}{999}
%
%
%
\bibitem{RefATLAS1}
R. Mackeprang; "Stable Heavy Hadrons in ATLAS"; Thesis, Niels Bohr Institute, University of Copenhagen, 2007.
\bibitem{RefATLAS2}
S. Tarem, S. Bressler, E. Duchovni, L. Levinson; "Can ATLAS avoid missing the long lived stau?"; ATL-PHYS-PUB-2005-022
\bibitem{RefATLAS3}
S. Hellman, M. Johansen, D. Milstead; "Mass Measurements of R-hadrons at ATLAS"; ATL-PHYS-PUB-2006-015.
\bibitem{RefCMS}
A. Rizzi; "Search and Reconstruction of Heavy Stable Charge Particles with the CMS experiment at LHC"; CMS TS-2007/016.
%
%
\end{thebibliography}
%

%
%
\end{document}